\documentclass{article}
\usepackage{graphicx}
\usepackage{calc}
\usepackage{epsfig}

\begin{document}
\title{Dynamics of learning in coupled oscillators
tutored with delayed reinforcements}

\author{M. A. Trevisan$^a$, S. Bouzat$^b$, I. Samengo$^b$, G. B. Mindlin $^a$\\
$^a$ Departamento de F\'{\i}sica, FCEyN, UBA, Argentina \\
$^b$ Centro At\'omico Bariloche, CNEA, Argentina
}
\date{\today}

\maketitle

\begin{abstract}
In this work we analyze the solutions of a simple system of
coupled phase oscillators in which the connectivity is learned
dynamically. The model is  inspired in the process of learning of
birdsong by oscine birds.  An oscillator acts as the generator of
a basic rhythm,  and drives slave oscillators which are responsible
for different motor actions. The driving signal
arrives to each driven  oscillator through two different
pathways. One of them is a {\em direct} pathway. The other one
is a {\em reinforcement} pathway, through which the signal arrives
delayed. The coupling coefficients between
the driving oscillator and the slave ones evolve in time
following a Hebbian-like rule.
We discuss the conditions under which a driven oscillator is
capable of learning to  lock to the driver. The resulting phase
difference and connectivity is a function of the delay of the reinforcement.
Around some specific delays, the system  is capable to generate dramatic
changes in the phase difference between the driver and the driven systems.

We discuss the dynamical mechanism responsible for
this effect, and possible applications of this learning scheme.
\end{abstract}

\section{Introduction}

Biological systems are capable of generating an extremely rich
variety of motor commands. Most impressively, in many cases,
these articulated commands are learned through experience. The
dynamical processes involved in learning are the focus of
extensive research, both in order to gain knowledge on how living
systems operate, as well as an inspiration for the design of
artificial systems capable of adaptation and learning.

In this framework, the acquisition of song by oscine birds is a
wonderful animal model for the study of how nontrivial behavior
can be learned \cite{brain00}, \cite{brain02}. First, it has
tight parallels with speech acquisition, since birds must hear a
tutor during a sensitive period, and practice while hearing
themselves, in order to learn to vocalize \cite{brain00},
\cite{brain02}. Second, the discovery of discrete nuclei (large
sets of neurons) involved in the process of producing and
learning song has provided a neural substrate for behavior,
turning this animal model in a rich test bench to study the
neural mechanisms of learning. Finally, recent physical models of
birdsong production have provided insight on how the activity of
different neural populations can be associated to acoustic
features of song \cite{mindlin04} \cite{prl}.

In the last years, much has been learned about the neural
processes involved in the generation of song by oscine birds. As
in many other biological systems, song production is based on a
rhythmic activity. Namely, a syllable is repeated when a motor
gesture is performed periodically. This gesture involves the
coordination of a respiratory pattern and the rhythmic activation
of the muscles controlling the syrinx (i.e. the avian vocal
organ). Recent work has unveiled that many important acoustic
features of birdsong are in fact determined by the phase
difference between two basic gestures: the  pressure at the air
sacs and the tension of the ventral muscles controlling the
syrinx \cite{prl}. As the result of an extensive research
program, the specific roles of different neural nuclei in this
process are being understood.

It was through lesions and observation of behavior that the
nuclei involved in the generation of motor activities responsible
for the production of song were identified \cite{nottebohm76}.
The so called  ``motor pathway'' is constituted by two nuclei
called HVC (high vocal center)  and RA (robustus nucleus of the
archistriatum). During the production of song, the HVC sends
instructions to the RA, which in turn, sends instructions to two
different  nuclei: the nXIIts, which enervates the syringeal
muscles, and the RAm, in control of
respiration.

Work on Zebra finches ({\em taeniopygia guttata}) has analyzed in
detail the exact time relation  between the firing of neurons in
HVC and RA during song \cite{hahnsloser02}. The picture emerging
from these experiments is that during each time window in the RA
sequence, RA neurons are driven by a subpopulation of
RA-projecting HVC neurons which are active only during that
window of time \cite{hahnsloser02}. These experiments suggest
that the premotor burst patterns in RA are basically driven by
the activities of HVC neurons (McCasland et al. 1981). If this is
the case, the architecture of the connectivity between the HVC
nucleus and the RA nucleus determines the complex patterns of
activity in RA neurons.

A second pathway thoroughly studied is the anterior forebrain
pathway (AFP). This pathway connects indirectly the nuclei HVC
and RA, as shown in Fig. 1.
\begin{figure}[htdf]
\begin{center}
\resizebox{6cm}{!}{\includegraphics{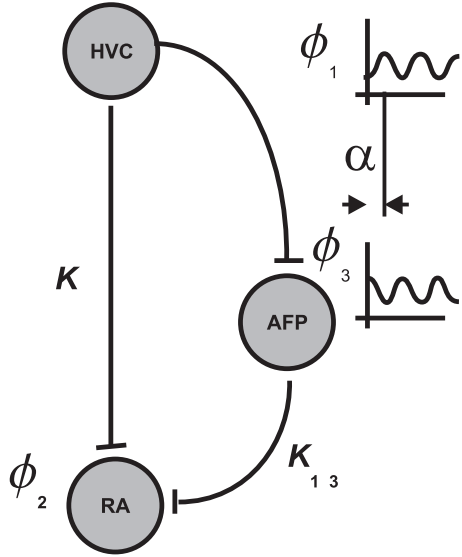}}
\end{center}
\caption{\label{f1} The activity of the nucleus HVC is
represented by the phase variable $\phi_1$, and drives the
neurons in nucleus RA through 2 different pathways: a {\it
direct} way synapses onto the cells described by the phase
variable $\phi_2$, and an {\it indirect} projection, controlling
the activity of $\phi_3$ of another set of cells. The phase
$\phi_3$ is delayed with respect to $\phi_1$ in $\alpha$. The
strength $k$ of the connection between $\phi_2$ and $\phi_1$
depends on the phase difference $\phi_2 - \phi_1$.}
\end{figure}
In contrast with the motor pathway, this pathway contributes only
minimally to the production of song in adults \cite{doupe2}.
However, it has been shown that the lesions to the nuclei in this
pathway during learning profoundly alters the bird's capability
of developing normal song \cite{botjer, nottebohm91}. The output
of this pathway is the lateral magnocellular nucleus of the
anterior neostriatum (LMAN). Individual RA neurons receive inputs
from both LMAN and HVC nucleus, which is consistent with the
picture that experience related LMAN activity facilitates certain
HVC-RA synapses, helping to build the neural architecture
necessary to produce the adult song. According to this picture, a
sequence of bursts generated at a RA-projecting HVC will induce
some activity in RA, and also eventually induce the activity of
LMAN that will lead to either the potentiation or depression of
the connection. This signal, however, requires a processing time
through the AFP, which has been estimated in approximately $40$
$ms$ in Zebra finches \cite{doupe2000}.

If we are interested in the generation of the instructions
controlling a motor output, it is sensible to model the process
in terms of global oscillators. In this way, the periodic
activity of the HVC nucleus when a syllable is being repeated is
represented in terms of the oscillation of a simple oscillator.
The nucleus RA, being driven by HVC, is represented by a second oscillator
driven by the first one. In this framework, the dynamics of
learning is the dynamics of the coupling coefficients between the
oscillators.

With this biological inspiration, we make a computational study of
the dynamical mechanisms by which a driven neural oscillator
(representing the activity of a subpopulation of RA neurons) can
learn to lock to its driver (representing the neurons in the
nucleus HVC). The forcing that HVC performs upon RA through the
indirect pathway AFP has been represented by a delayed
reinforcement. The learned phase difference between driver and
driven oscillators has been studied as a function of the
reinforcement delay. In certain parameter regimes, small changes
in the delay have been found to lead to important changes in the
learned phase difference, and we explain this effect in terms of
the dynamics of the system.

The work is organized as follows. In section 2 we discuss the
mathematical model used to emulate the activities in our neural
circuits. The solutions of this model are discussed in section 3.
The consequences of the dynamical skeleton in terms of learning
dynamics are discussed in section 4. Section 5 contains applications
of these mechanisms to rate models of neural populations. Section 6
presents our conclusions.

\section{Model}

Rhythmic activity play an important role in many neural systems
\cite{lev}. Cyclic neuronal activity is typically modeled in terms
of periodic oscillators. Moreover, for those cases where the
amplitude of the oscillations does not vary, it is possible to
further reduce the dynamics of the oscillator to a phase
variable. Winfree, Kuramoto and others have shown important
features of coupled systems following this approach \cite{kurths}.

In a recent work, a generalized Kuramoto model of coupled
oscillators with a slow coupling dynamics was investigated
\cite{lev}. Inspired by a Hebbian like learning paradigm, the
authors wrote a dynamical system for the evolution of the
coupling which would strengthen synchronized states. The system
of oscillators presented an interesting dynamics: the original
difference between the natural frequencies of the oscillators
served as a driving force in the dynamics of the phase
differences between the oscillators. When the coupling parameters
fell within a given range, the oscillators were eventually able
to lock. In addition, a plasticity ingredient was incorporated,
acting at a slower time scale. Namely, the dynamics of the
coupling strength between the oscillators was driven by the phase
difference between them. Once again, for some parameter values,
the oscillators would end up locked.

In this work we are interested in the process of locking an
oscillator to its driver, exploring the dynamical
consequences of reinforcing the driving through a second pathway.
In order to model this effect, three phase variables $\phi_1$,
$\phi_2$ and $\phi_3$ are introduced. In terms of the inspiring
problem, $\phi_1$ represents the oscillation of a first nucleus
generating a periodic instruction (as HVC, with its periodic dynamics).
The phase $\phi_2$
stands for the activity of some region of the nucleus RA, which
contains premotor neurons controlling some aspect of the song
production. This phase is driven by $\phi_1$. Finally, $\phi_3$
parametrizes the activity of the indirect pathway, and its
dynamics is assumed to be the same as that of $\phi_1$,
delayed some time $\tau$. Modelling  the
activity of HVC as a simple a harmonic function of frequency
$\omega_1$, this delay translates into a phase
$\alpha=\omega_1 \tau$.

According to these hypothesis, the model describing the dynamics
of these variables reads:

\begin{eqnarray}
\frac{d\phi_1}{dt} & = &\omega_1 \\
\frac{d\phi_2}{dt} & = &\omega_2-k \sin(\phi_2-\phi_1)-k_{13}
\sin(\phi_2-\phi_3) \\
\frac{dk}{dt} & = & \gamma \cos(\phi_2-\phi_1) - k
\end{eqnarray}

\noindent with $\phi_3=\phi_1-\alpha$.

\noindent In Fig. 1 we show a sketch of the three nuclei and
their connections indicating the associated dynamical variables.
Replacing $\phi_3=\phi_1-\alpha$ in Eq. 2, and scaling the
equations, the following dynamical system for the phase difference
$\phi_2-\phi_1=\phi$ and $k$ is obtained,

\begin{eqnarray}
\frac{d\phi}{dt} & = & 1- k \sin\phi - k_{13} \sin(\phi + \alpha) \\
\frac{dk}{dt} & = & \epsilon (\gamma \cos\phi - k).
\end{eqnarray}

\noindent We are interested in understanding the dynamics of this
set of equations, stressing on the stationary solutions that the
system can ``learn''. We are particularly interested in finding
out whether there are delays that can provide any special
advantage in the process of learning to control a periodic motor
pattern.

\section{Solutions}

For a qualitative understanding of the solutions presented by
this system of equations, we can compute the nullclines: curves
for which each variable is stationary. They are:

\begin{eqnarray}
k & = & [1 - k_{13}\sin(\phi + \alpha)] / \sin\phi.
\label{nullcline2}\\
k & = & \gamma \cos(\phi) \label{nullcline1}
\end{eqnarray}

\noindent In Fig. 2 we display the nullclines for different values
\begin{figure}[htdf]
\begin{center}
\resizebox{6cm}{!}{\includegraphics{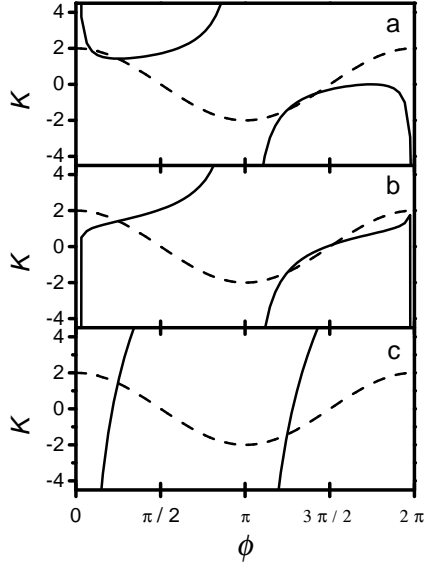}}
\end{center}
\caption{\label{f2} Nullclines $\dot{\phi} = 0$ (full line) and
$\dot{k}= 0$ (dashed line), $\gamma=2$, $\alpha=3\pi/4$ and
different values of $k_{13}$: (a) $k_{13}= 1.5$, (b) $k_{13} =1$,
and (c) $k_{13} = 10$.}
\end{figure}
of the parameters. The intersections of the nullclines give the
fixed points of the system $(k_0, \phi_0)$.

In the parameter range $k_{13} \in [0, 1.42]$,  the nullcline of Eq.
(\ref{nullcline2}) presents two branches: the first one with a
minimum, the second one with a maximum. Depending on the
parameter values, one of the branches or both might intersect the
nullcline of Eq.(\ref{nullcline1}). These intersections, when they occur,
lead to the appearance of a saddle and a node in a saddle node
bifurcation. For $\gamma$ sufficiently large, two attracting
fixed points (separated by two saddles) can coexist. On the other
hand, for $\gamma$ sufficiently small, there are no
intersections between the nullclines and therefore, no stationary
phase difference between the driver and the driven oscillator can
be established.

The topological organization of the nullclines present
qualitative changes as the system parameters are varied. These
changes leave their imprint in the dependence of the stable fixed
point angular positions with the parameters. In Fig. 3 we show the
\begin{figure}[htdf]
\begin{center}
\resizebox{10cm}{!}{\includegraphics{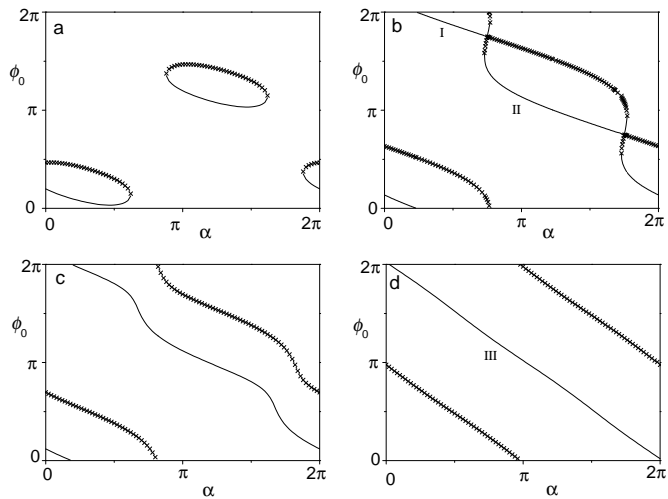}}
\end{center}
\caption{\label{f3} Stationary solutions for $\phi$ as a function
of $\alpha$ for $\gamma=1$. Results for (a) $k_{13}=0.9$, (b)
$k_{13}=1.5$, (c) $k_{13}=1.8$ and (d) $k_{13}=15$. In all the
cases the solid lines correspond to stable solutions and the
crosses indicate the branches of unstable solutions.}
\end{figure}
positions of the  fixed points as a function of the reinforcement
delay $\alpha$, varied between $(0, 2 \pi)$. The different insets
correspond to a different value of the reinforcement parameter
$k_{13}$. In the figure, the solid lines indicate the linearly
stable solutions and the lines with crosses the unstable ones
(which in all cases are saddle points). For small values of the
coupling, there are delays for which no fixed points exist (Fig.
3a). At specific delays, stable and unstable fixed points are
born in saddle node bifurcations. In terms of nullclines, this
corresponds to situations in which the second nullcline of
Eq.(\ref{nullcline2}) presents a minimum which touches the
nullcline of Eq. (\ref{nullcline1}) (see Fig 2.a). As the
reinforcement strength is increased, the region of reinforcement
delays for which no stationary solutions exist decreases in size.
As the reinforcement parameter $k_{13}$ is further increased, the
regions with no solutions disappear and the bifurcation curves
meet at a transcritic point (Fig. 3b) around which, a narrow zone
of multistability appears. Finally, as the reinforcement strength
is further increased,  the angular position of the fixed points
varies monotonically (Figs. 3c, 3d).

The existence of these bifurcation curves has profound
consequences in terms of learning. Notice that close to certain
values of the delay $\alpha$, a minimal change in $\alpha$ gives
rise to important differences in the equilibrium phases learned by
the system. In the following section, we will discuss potential
consequences of this bifurcating structure in a learning process.

\section{Interpretation}

The animal model inspiring our dynamical model is the motor
pathway in oscine birds. Part of this pathway is the nucleus RA
containing excitatory neurons, some of which enervate respiratory
nuclei, and others enervate the nucleus nXIIts, that projects to
the muscles in the syrinx \cite{spiro}. These two populations are
segregated into different regions of the RA structure.

Recently, the study of the avian vocal organ allowed us to
associate acoustical features of the song with properties of the
muscle instructions necessary to generate the song. The production
of repetitive syllables requires a cyclic expiratory gesture, and
a cyclic gesture of the syringeal muscles \cite{gardner}. Sound is
produced by labia located at the junction between bronchii and
tract, obstructing periodically the airflow . The model mentioned
above describes the departure of the midpoint of a labium from
the prephonatory position, $x$ \cite{prl,pre}:

\begin{eqnarray}
\frac{dx}{dt} & = & y  \label{std2eq1}\\
\frac{dy}{dt} & = & -\epsilon(t) x- C x^2 y + B(t) y \label{std2eq2}
\end{eqnarray}

\noindent where $\epsilon(t)$ is a function of the activity of
ventral muscles, whereas $B(t)$ is a function of the bronchial
pressure. This model has been tested, by using experimentally
recorded $\epsilon(t)$ and $B(t)$. The resulting $x(t)$ was
remarkably similar to the one produced while the physiological
data had been recorded \cite{pre}.

The phase difference between the $\epsilon(t)$ and $B(t)$,
responsible for important acoustic features of song (such as the
dynamics of the syllabic fundamental frequency) originates in RA.
Recent work has unveiled that direct connections between
respiratory nuclei and nXIIts can affect the final value of the
phase difference. Yet, it is at RA that the neurons driven by HVC
also receive input from the indirect pathway AFP, and therefore
it is at this level that the phase difference between gestures
can be altered.

In order to apply the results of the previous section to a
learning strategy for birdsong, let us assume that two oscillators
represent the cyclic activity of RA during the production of song.
One of the oscillators mimics the activity of the neural
population enervating the nXIIts nucleus, while the other
oscillator represents the population of neurons which control the
respiratory pattern. Both oscillators are assumed to be driven by
the nucleus HVC, presenting a global activity with a syllabic
rhythm. The reinforcement oscillator drives both RA oscillators
with a signal equal to that of HVC, but delayed in a phase
$\alpha$. Figure 4a illustrates the proposed architecture.
\begin{figure}[htdf]
\begin{center}
\resizebox{5cm}{!}{\includegraphics{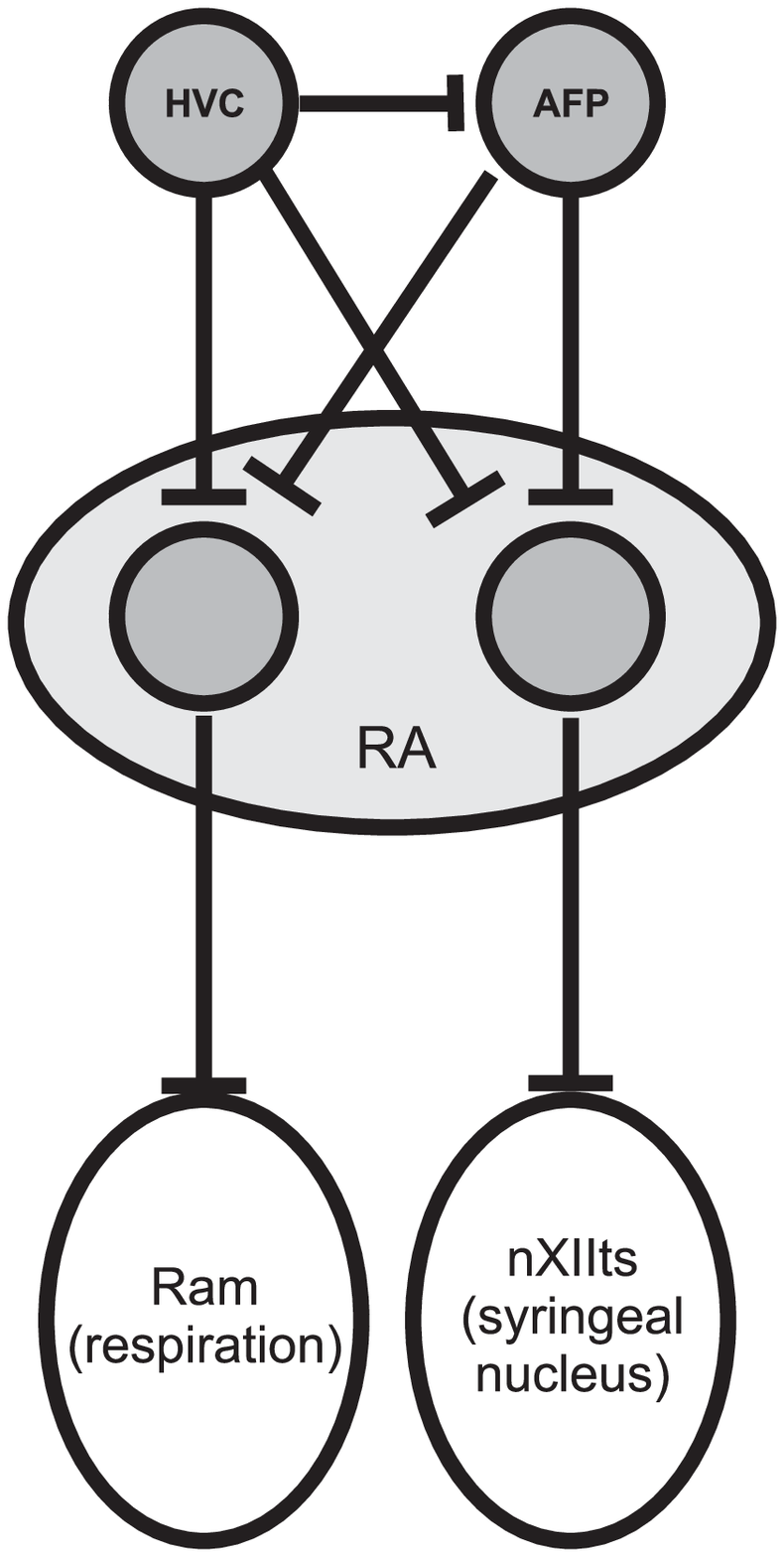}} \hspace*{1cm}
\resizebox{6cm}{!}{\includegraphics{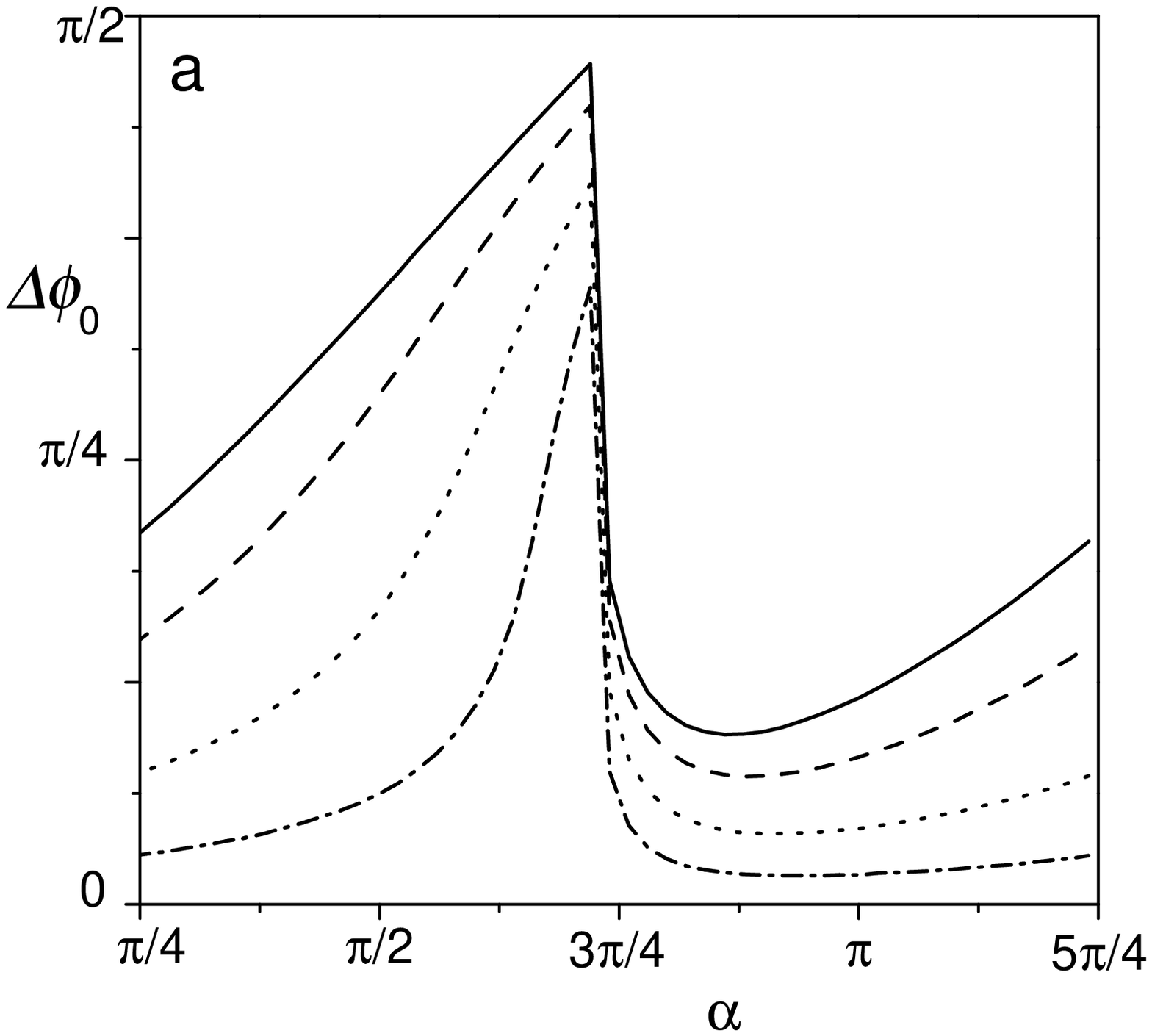}}
\end{center}
\caption{\label{f4} (a) Two oscillators representing neuron
populations (nXIIts and RAm) driven by a master oscillator (HVC)
will learn to follow the master at different delays. The two
populations control different aspects of the song production
apparatus: the syringeal nucleus and the respiratory muscles,
respectively. (b) Phase difference $\Delta \phi_0$ as a function
of $\alpha$. The solid line corresponds to the difference between
the stationary solutions with $k_{13}=15$ and $k_{13}=1.5$; the
dashed line to $k_{13}=5$ and $k_{13}=1.5$; the dotted line to
$k_{13}=2.5$ and $k_{13}=1.5$; and the dashed-dotted line to
$k_{13}=1.8$ and $k_{13}=1.5$. All calculations are performed
with $\gamma=1$.}
\end{figure}

According to Fig. 3, if the two oscillators representing RA are
strongly reinforced by the AFP circuit (i.e., if $k_{13}$ is
large as in Fig. 3d), for every value of the forcing there will
be a locked state. The phase difference between their own
oscillation and the one of the driver will be the same for both
oscillators, whatever the value of the delay $\alpha$. However, a
different situation is found if each one of the oscillators is
reinforced through a different coupling strength $k_{13}$.
Imagine that for one of them, $k_{13}$ is similar to the one used
to generate Fig. 3b, while the other one is forced though a
coupling strength as the one used to generate Fig. 3d. In this
case, depending on the value of the delay, qualitatively
different phase differences can be achieved between each RA
oscillators and the driver (and therefore, between the two RA
oscillators themselves).

In Fig. 4b, the value of the phase difference $\Delta \phi_0$ is
displayed as function of $\alpha$ for different pairs of
oscillators. We have fixed $\gamma=1$. In all cases, one of the
oscillators, which is taken as a reference, is assumed to be
coupled to the AFP circuit through the parameter $k_{13}=1.5$,
corresponding to the situation of Fig. 3b. For the other
oscillator, we consider different values of $k_{13}>1.5$. All
these couplings give rise to stationary solutions $\phi$ that
decrease monotonously as a function of $\alpha$ (see Figs. 3c and
3d). The curves in Fig. 4b are the difference between the
stationary solution of both oscillators, for several values of the
second coupling constant. In order to indicate in detail how this
difference is computed we take the example of the solid line in
Fig. 4b, for which the second oscillator is coupled to the AFP
circuit with a strength $k_{13}=15$. This curve corresponds to
the difference of the solutions indicated with I and II in Fig.
3b (i.e. the stationary solution for the reference oscillator)
and the curve III on Fig. 3d (the stable solution for the second
oscillator). We notice, however, that there is a small range of
$\alpha$ for which the branches of solutions I and II coexist.
For such values of $\alpha$ we have taken branch II as the
solution for the reference oscillator. This choice leads to the
lowest possible value of phase difference between the two
oscillators, since branch III is closer to branch II than to
branch I. (Hence, the phase difference could be even higher than
the one shown in Fig. 4b). It should be stressed that there are
delays (close to the value $\alpha=3\pi/4$ for the parameters
used here) for which small displacements can lead to a huge
change in the stationary phase difference between the
oscillators. Notice that Fig. 4b is $\pi$-periodic, so there is a
similar effect around $\alpha = 7 \pi / 4$.

The results in Fig. 4b, are robust with respect to changes in the
parameter $\gamma$. We have checked that for all $\gamma \in
(0.5, 10)$, the transcritical bifurcation occurs at $\alpha
\approx 3\pi/4$ and $\alpha \approx  7 \pi / 4$. Moreover,
whenever $\alpha$ is in the neighborhood of any of these critical
values, the phase difference between oscillators with different
$k_{13}$ is highly sensitive to the value of $\alpha$. Yet, the
values of $k_{13}$ relevant for observing the mentioned phenomena
do depend on $\gamma$. For instance, the value of $k_{13}$ at
which the transcritical bifurcation is observed increases with
$\gamma$. It goes from $k_{13}\simeq 1.2$ for $\gamma=0.5$ to
$k_{13}\simeq 6$ for $\gamma=10$.

Figure 5 illustrates how different the learned syllables can be
\begin{figure}[htdf]
\begin{center}
\resizebox{6cm}{!}{\includegraphics{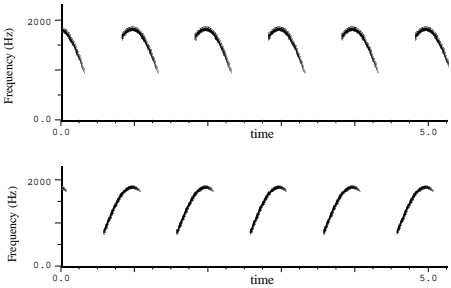}}
\end{center}
\caption{\label{f5} The sonograms of two songs which consist of a
repetition of the same syllable, for which the phase difference
learned is (a) $\Delta \phi_0=0$ and (b)  $\Delta \phi_0=\pi/2$.
The parameters of Eqs. \ref{std2eq1} and \ref{std2eq2} are
$\epsilon(t)=7\, 10^7 + 6\, 10^7 cos(2\pi t/44100)$ a.u,
$B(t)=5\,10^2+10^3 cos(2\pi t/44100+\Delta \phi_0)$ a.u, $C=2 \,
10^9$ a.u.}
\end{figure}
for small changes in the reinforcement delay, if they occur
around $\alpha=3\pi/4$. The figure displays a sonogram showing
the time evolution of the fundamental frequency of the sound
produced by the model of the syrinx, when driven by the
``learned'' phase differences.

    \section{Application to rate models}

The Kuramoto model describing the time evolution of phase
differences between oscillators constitutes a popular model,
particularly suited for analytic work. Yet, we explored whether these
effects are
also present
in other models. We tested the basic findings of the previous
sections in rate models for the activities of neural populations.

Rate models are introduced to account for the dynamics of the
average activity of neural nuclei. For a problem involving a
macroscopic motor control program it is a suitable level of
description. In particular, a widely used model for the average
activity of two subpopulations (exitatory and inhibitory) is the
Wilson Cowan system of equations \cite{izhikevich}.

We let $x$ and $y$ stand for the activities of  the excitatory and inhibitory
subpopulations respectively. In \cite{spiro} it was shown that in the portions
of RA with neurons projecting to XIIts and to respiratory nuclei, the
(excitatory) projecting neurons coexist with inhibitory neurons.
 We drove the equations ruling their dynamics
 with two signals.  One
represents HVC activity and the other one, the activity in nucleus
LMAN, assumed to be a delayed copy of the first signal. A typical
hebbian rule was used to describe the dynamics of the coupling
between HVC and RA. The system reads

\begin{eqnarray}
\frac{dx}{dt} & = & -x + S[\rho_x + ax + by + k\cos(\omega t) +
    k_{13}\cos(\omega t - \alpha)] \nonumber \\
\frac{dy}{dt} & = & -y + S(\rho_y + cx + dy)  \\
\label{wcowan} \frac{dk}{dt} & = & \lambda x\cos(\omega t) - k
\nonumber
\end{eqnarray}

with a saturation function $S(x)=1/(1+e^{-x})$.

We integrated these equations for the whole range of delays
$\alpha\in[0,2\pi)$. In Fig. 6 we show a window representing a
\begin{figure}[htdf]
\begin{center}
\resizebox{8cm}{!}{\includegraphics{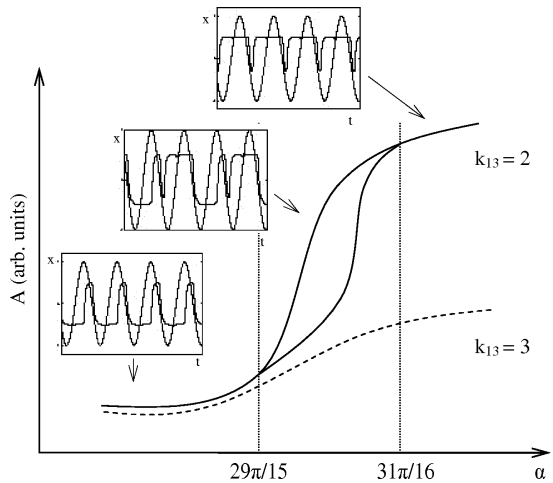}}
\end{center}
\caption{\label{f6} Bifurcation diagram for the Wilson-Cowan
system (Eq. 10) forced with frequency $\omega = 0.3$. The
horizontal axis represents the delay $\alpha$ and the vertical
axis represents the maximum amplitude $A$ of the oscillations in
phase space. A broad phase difference between two slave
oscillators with weights $k_{13}=2$ and $k_{13}=3$ can be
achieved within a narrow range of delays
$29\pi/15<\alpha<31\pi/16$. On the insets, solutions $x(t)$ along
with forcing function $cos(\omega t)$ for the three different
regimes found. The equations were integrated with the following
parameters: $\rho_x=-5.75$, $\rho_y=-1$, $a=10$, $b=-1.5$, $c=2$,
$d=2$ and $\lambda=68$.}
\end{figure}
bifurcation around $\alpha=29\pi/15$, where two qualitatively
different period-one solutions can be found. The transition
occurs through a bubble in parameter space where a period-two
exists. It is interesting that the different period-one solutions
(illustrated in the insets), are locked to the periodic forcing
frequency at different phases. As $k_{13}$ is increased, only a
single period-one solution is present. Therefore, two
subpopulations reinforced through a delay $\alpha$ around this
bifurcating value with these weights will lock at a phase
difference within a wide range of values as the delay is slightly
increased.

\section{Conclusions}

In a recent work \cite{lev}, the process of learning a phase
difference between two phase oscillators coupled with a slow
varying coupling constant was described. Here, the impact of
subjecting the driven system to two delayed copies of a signal is
studied. This design is inspired in an animal model, oscine birds,
which learn their song by modifying the architecture of
connections within the motor pathway. Our model is a caricature
of the complex processes that have been described for the animal
model, which involved the convergence to a motor nucleus of two
signals, separated by a delay. Our study indicates that for some
values of the amplitude of the reinforcement, the learned phase
difference between driver and slave  depends strongly on the delay
of the reinforcement. This allows to conceive simple strategies of
learning motor gestures that require a tuning between different
neural populations. We precisely illustrate the power of this
strategy with an example of the inspiring biological problem. We
show that playing with small variations in the delay of the
reinforcement, completely different syllables in a bird song can
be generated.

In the problem of bird song learning, the timing of the indirect
pathway (the anterior forebrain pathway) can be changed
\cite{abarbanel04} by dopaminergic input. Not much is known about
the precise nature of the coding used by a bird to translate an
error into the indirect signal. Measures of activity in the
nucleus LMAN (the last one of the AFP pathway, which projects
onto RA) in a juvenile bird learning his song are still not
possible. The simple model presented in this work allows us to
explore theoretically the dynamical mechanisms that enter into a
learning scheme compatible with the basic ingredients present in
the animal model. Moreover, it provides a new control mechanism
applicable to artificially designed neural systems.

\section*{Acknowledgements} This work was partially funded by
UBA, CONICET, ANPCyT, Fundaci\'on Antorchas and NIH.

\end{document}